\documentclass[pra,aps,twocolumn,superscriptaddress,showpacs]{revtex4}
\usepackage{graphicx}
\usepackage{dcolumn}
\usepackage{amssymb}

\begin{document}
\title{Quantum algorithms for phase space tomography}

\author{Juan Pablo \surname{Paz}}
\email{jpaz@lanl.gov}
\affiliation{Departamento de F\'{\i}sica, FCEyN, UBA, Pabell\'on 1,
Ciudad Universitaria, 1428 Buenos Aires,
 Argentina}
\affiliation{Theoretical Division, LANL, MSB213, Los Alamos, NM 87545, USA}

\author{Augusto Jos\'e \surname{Roncaglia}}
\email{augusto@lanl.gov}
\affiliation{Departamento de F\'{\i}sica, FCEyN, UBA, Pabell\'on 1,
Ciudad Universitaria, 1428 Buenos Aires, 
Argentina}
\affiliation{Theoretical Division, LANL, MSB213, Los Alamos, NM 87545, USA}

\author{Marcos \surname{Saraceno}}
\email{saraceno@tandar.cnea.gov.ar}
\affiliation{Unidad de Actividad F\'{\i}sica, Tandar, CNEA, Buenos Aires, Argentina}

\date{\today}

\begin{abstract}
We present efficient circuits that can be used for the phase space tomography of quantum
states.  The circuits evaluate individual values or selected averages of the Wigner, Kirkwood
and Husimi distributions. These quantum gate arrays can be programmed by initializing appropriate 
computational states.  The Husimi circuit relies on a subroutine that is also interesting in 
its own right: the efficient preparation of a coherent state, which is the ground state of the 
Harper Hamiltonian.  
\end{abstract}


\maketitle

\section{Introduction}
 
Phase space distributions have been used as representation tools for quantum mechanical
operators since the early days of quantum mechanics.  They provide the ideal link to explore
and understand the transition to classical mechanics and to display in phase space quantum
effects.  Their properties are very well known \cite{Wigner,Balazs} when the phase space is 
$\mathbb{R}^2$. For systems with a finite dimensional space of states the distributions become discrete, i.e.
they are defined over a finite lattice \cite{Wootters, Leonhardt}. Discrete phase space distributions
have been used in the context of studies of quantum maps on bounded phase space \cite{Ozorio,Berry}
and they have also recently proposed as a useful tool for studies related to quantum information
and computation \cite{MPSpra,Paz,Buzek}. The simplest way to characterize them, for a 
Hilbert space of dim $N$ is by using a complete basis of $N^2$ operators 
$\{ \Lambda_\alpha ; \alpha = 0,..N^2 -1\}$, in terms of which the distribution is given as
$\text{Tr} [\Lambda_{\alpha}^\dagger \rho]$. The properties and classical features
that these distributions display depend of course on the operator basis $\{ \Lambda_\alpha \}$.
In this sense phase space distributions are nothing but the coefficients of the expansion of the
state $\rho$ in the basis $\Lambda_\alpha$.  The determination of the value of $\text{Tr}[\Lambda_\alpha \rho]$
for every $\alpha$ is, thus, a particular form of quantum state tomography (see \cite{Dariano} and references
therein).

In a recent paper \cite{MPSnat} it was shown how to efficiently measure the discrete Wigner function
at any phase space point.  The basis of the method is the use of the so-called 'scattering circuit'
to efficiently determine the value of the quantity $\text{Tr}[\rho A]$ provided that the operation
$A$ can be implemented in a controlled way.  Thus, if the complete basis $\Lambda_\alpha$ consists
of unitary operators then the scattering circuit can be used to measure individual values
of the distribution.  The disadvantage is of this approach is that, as $\alpha$ enters as a 
classical parameter, a new gate array has to be applied for each $\alpha$.  In this paper we 
will extend the results presented in \cite{MPSpra,MPSnat} in two ways.  
First, we will show how to efficiently measure other phase space distribution functions 
(Husimi, Kirkwood). Second, we will show how to do this by using quantum circuits with a fixed architecture,
which is independent of the phase space point $\alpha$.  These circuits belong
to the class of programmable quantum devices, whose action is controlled
by quantum software, that have been under investigation recently \cite{Qprogram,PR}. 
In this paper we will show how to build efficient programmable circuits to measure three phase 
space distributions:  Wigner, Kirkwood and Husimi.  It is also worth mentioning 
here that the quantum circuits we developed use a subroutine which is interesting in its own 
right and could be useful for other applications. In fact, in this paper we present 
a method to efficiently prepare coherent sates (which are rigorously defined below, but can 
be roughly characterized as approximately Gaussian wave packets obeying periodic boundary conditions).

The paper is organized as follows in Section \ref{Wig} we present the circuit
that enables the programmable measurement of the discrete Wigner function. We also show that
it can be useful to compute averages of this function over various phase space domains (this 
extends and completes results presented in \cite{PR}). In Section \ref{Kirk} we present 
a simple programmable circuit that evaluates the Kirkwood distribution at any phase 
space point. In Section \ref{husprog} we present the quantum gate array that efficiently 
evaluates the discrete Husimi distribution. This gate uses coherent states as inputs. The 
algorithm to efficiently prepare those states is presented in Section \ref{Coh}. Finally, 
we present some conclusions in Section \ref{Concl}.

\section{Programmable tomography of the discrete Wigner function} \label{Wig}
 
The discrete Wigner function \cite{MPSnat,MPSpra} in a Hilbert space of dimension $N$ is defined 
in terms of the basis of phase point unitary operators:
\begin{equation}
A(q,p)=U^q R V^{-p} e^{i\frac{\pi}{N}pq},\label{ppoperator}
\end{equation}
as
\begin{equation}
W(q,p)=\frac{1}{2N}\text{Tr}[A(q,p)\rho]
\end{equation}
where $q,p$  are integer labels spanning a grid of size $2N\times 2N$. $U$ and $V$ are respectively
the translation operators in the $|q\rangle$ and $|p\rangle$ basis 
($U|q\rangle=|q+1\rangle$, $U|p\rangle=e^{-i\frac{2\pi}{N}p}|p\rangle$,
$V|p\rangle=|p+1\rangle$,$V|q\rangle=e^{i\frac{2\pi}{N}q}|q\rangle$), which are related
by the discrete Fourier transform. $R$ is the reflection operator ($R|n\rangle=|N-n\rangle$). Only an
$N\times N$ sub-grid is needed for the complete tomography of the state (but the larger grid is
required to define a Wigner function with all the desired properties \cite{MPSpra}).  
 
\begin{figure}[ht]
\includegraphics[width=8cm]{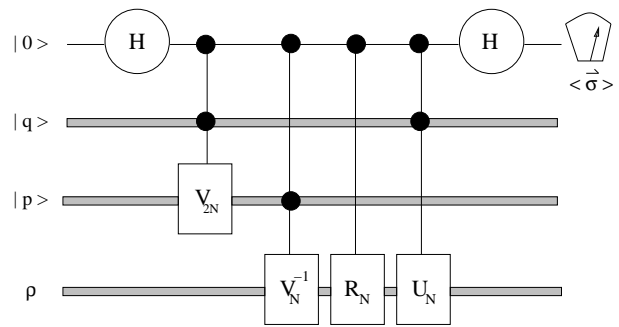}
\caption{Programmable circuit to evaluate the discrete Wigner function from the
polarization of the first qubit. The second and the third registers 
store the information about the the phase space point where the distribution is being
measured. All states are initialized in the computational (coordinate) basis. 
All ``ctrl-$O$'' operations act as 
(ctrl-$O$)$|n\rangle|\Psi\rangle=|n\rangle O^n|\Psi\rangle$. A subscript in an operator 
denotes the dimension of the space in which it acts. In all figures we  adopt the following graphic
convention: thin cables denote a single qubit, and wide cables denote systems of more than one qubit.}\label{wigner}
\end{figure}

The programmable circuit implementing the measurement of the discrete
Wigner function is shown in Figure \ref{wigner}. This was introduced in \cite{PR} as a particular
case of a programmable circuit evaluating the expectation value of an arbitrary operator. It is
a variation of the so--called scattering circuit \cite{MPSnat} where an ancillary qubit 
acts as a probe for a more complex system with which it interacts by means of controlled 
operations. The circuit shown in Figure \ref{wigner} has several registers: The first register
is an ancillary qubit (the probe) initially prepared in the state $|0\rangle$ which is an eigenstate
of $\sigma_z$ with eigenvalue +1. The following
two registers act as program registers and should be prepared in the state $|q\rangle |p\rangle$.
The last register stores the state of the system of interest $\rho$. The program state contains
the information about the binary expansion of the coordinates of the point in the phase space 
where we wish evaluate the Wigner function. As seen in the circuit, the role of the 
program states is to control the application of displacement operators on the system register. 
Here, and in what follows, we use the convention that for any operator $O$, 
``controlled-$O$'' operators act as: (ctrl-$O$)$|n\rangle|\Psi\rangle=|n\rangle O^n|\Psi\rangle$.
In particular, in Figure \ref{wigner} an operator such as ``control-$V_{2N}$'' acts 
as (ctrl-$V_{2N}$)$|q\rangle|p\rangle=|q\rangle V_{2N}^{q}|p\rangle=e^{i\frac{2\pi}{N}pq} 
|q\rangle|p\rangle$ (note that a subscript in any operator indicates 
the dimensionality of the space in which it acts).  
It is straightforward to show that the final polarization of the ancillary qubit turns out to be: 
\begin{equation}
\langle \sigma_z \rangle = 2N\text{Tr}[A(q,p)\rho]=2NW(q,p).\label{measurewig}
\end{equation}

\subsection{Measuring the sum of the discrete Wigner function over domains in phase space}

One of the defining properties of  the Wigner function is the fact that adding 
its values over lines in phase space one always obtains the probability to measure an observable. 
It is interesting to notice that the circuit shown in Figure \ref{wigner} can be programmed
to directly evaluate the average of the Wigner function along any line in phase space. 
More generally, the state of the program register can be used to define the phase space 
domain over which the Wigner function is averaged.

Let us consider first the case of lines. The quantity in which we are interested is 
$\sum_{(q,p)\in L}W(q,p)$, the sum of the values of the Wigner function along the line $L$. 
It is easy to see that for the program state 
$|\Psi\rangle_P = \sum_{p=0}^{2N-1}|n_3\rangle\otimes|p\rangle/\sqrt{N}$, 
the final polarization turns out to be $\langle \sigma_z \rangle = \sum_p W(n_3,p)$. 
As this type of program state can be efficiently constructed, the example shows that it 
is possible to estimate the sum of the values of the Wigner function along vertical and horizontal
lines. In a recent work \cite{PR} we showed that this is a special case of a more general
result that establishes the possibility to program the measurement of the expectation value
of any operator. Following the same idea, consider the program state 
\begin{equation}
|\Psi\rangle_P =\frac{1}{K} \sum_{(q',p') \in L} |q'\rangle|p'\rangle
\end{equation}
where $L: n_1q+n_2p=n_3$, $0\leq n_i\leq 2N-1$ and $K$ is a normalization constant (the square
root of the number of points in the line $L$). Then, the final polarization is:
\begin{equation}
\langle \sigma_z\rangle=\frac{2N}{K^2}\sum_{(q',p') \in L}W(q',p').
\end{equation}
This is precisely the quantity we are interested in. However, the above program states may be 
difficult to prepare (they are, in general, highly entangled states). To avoid using 
program states which may be difficult to prepare we have developed an alternative method. This
was briefly described in \cite{PR}. For completeness, we present it here in more detail. 

\begin{figure}[ht]
\includegraphics[width=8cm]{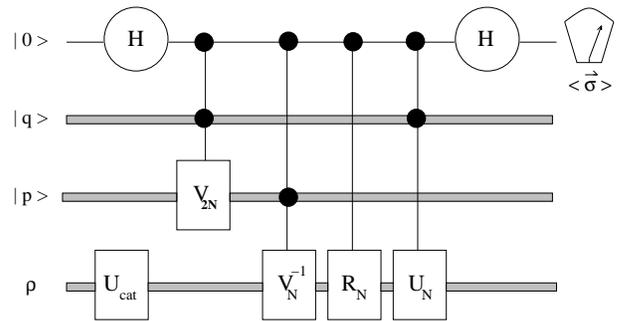}
\caption{Programmable gate array to evaluate the average Wigner function over
a tilted line in phase space. The cat map is parameterized by two integers, which can be 
programmed in auxiliary registers (so the array can be made fully programmable).}\label{sum}
\end{figure}

For our method it is convenient to employ the fact that certain unitary operators 
induce a purely classical transformation of the Wigner function (this means that 
the Wigner function is simply transported by a canonical, area preserving, flow).
This is the case for unitary operators that quantize linear canonical transformations 
on the torus (the so called cat maps \cite{MPSpra}).  We can use this fact as follows: 
First, we can prepare a simple program state that would produce the measurement of the
Wigner function along vertical or horizontal lines. Then, we can obtain the corresponding 
measurement along tilted lines by applying the appropriate unitary cat map to the initial 
state. The two-parameter family of cat operators that we will use is given by:
\begin{equation}
U_{cat}={\cal V}_b{\cal T} {\cal V}_a, 
\end{equation}
where $a$ and $b$ are integers, and the operators ${\cal V}_a$ and $\cal T$ are diagonal
in the position and momentum basis respectively,
\begin{eqnarray}
{\cal V}_a |n\rangle & = & \exp(-i2\pi n^2 (1-a)/2N)|n\rangle\nonumber\\
{\cal T} |k\rangle & = & \exp(-i2\pi  k^2/2N)|k\rangle.
\label{kicks}
\end{eqnarray}
The classical equations of motion corresponding to this system are
\begin{equation}
q= b q' + p'\qquad p= (ab-1)q' + ap'.\label{classicalcat}
\end{equation}
As the Wigner function evolves classically, when the phase space points are related 
as above, we can write $W(q,p,t+1)=W(q',p',t)$. 
The transformation (\ref{classicalcat}) maps vertical lines into tilted lines
according to the values of the parameters $a$ and $b$. With this in mind we can try to 
find the linear transformation that maps a line $L$, 
whose  program  state is difficult to prepare, into a line $L'$, 
whose program state is easy to prepare. If we achieve this, we can compute the average
Wigner function along $L$ by using the fact that
\begin{displaymath}
\sum_{(q,p) \in L}W_{\rho}(q,p)=\sum_{(q',p')\in L'}
W_{U_{cat}\rho U_{cat}^{\dag}}(q',p'). 
\end{displaymath}
The circuit to implement this procedure is shown in Figure \ref{sum}. In order to implement the $U_{cat}$
evolution, we only need to know how to apply the unitary operator ${\cal T}$ and its powers.
As it was shown in \cite{Shepelyansky} this can be done efficiently.   
Taking into account that the parameters defining the evolution $U_{cat}$ depend on the line,
this network is not completely programmable (its architecture depends on each
line), but we can easily prove that the device can transformed into a fully programmable 
one by adding two registers specifying the values of the constants $a$ and $b$. 

Let us now address the issue of how to find the parameters $a$ and $b$ entering 
in (\ref{classicalcat}) mapping line $L$ and $L'$. For this, we consider the lines
defined as 
\begin{eqnarray}
L  &:& n_1q+n_2p=n_3 \quad \text{mod 2N} \nonumber\\
L' &:& q'+p'=n_3 \quad \text{mod 2N}. \nonumber
\end{eqnarray}
For simplicity, we consider the case where at least one of the parameters 
$n_i$ is an odd number (the other case can be treated similarly). It is worth 
mentioning that the program state for $L'$ can be efficiently prepared. 
The mapping between the two lines is accomplished by using a cat map as 
in (\ref{classicalcat}) with the parameters given by
\begin{eqnarray}
n_2a &=& 1-n_1 \quad \text{mod 2N} \nonumber \\
b &=& 1+n_2 \quad \text{mod 2N}.\quad
\end{eqnarray}

This method can be generalized to evaluate the average value of the Wigner function 
over tilted rectangular regions. For this purpose, we can construct the program 
state for a simple rectangular region (defined by the conditions: 
$q_1\leq q \leq q_2$, $p_1\leq p \leq p_2$). Then, we can map this region into 
a tilted region by using the strategy described above. 

\section{Programmable tomography of the Kirkwood distribution} \label{Kirk}

\begin{figure}[ht]
\includegraphics[width=8cm]{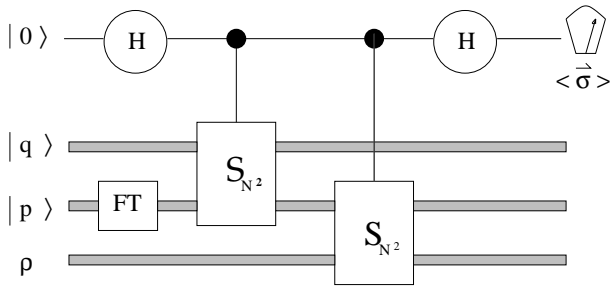}
\caption{Programmable gate array to  measure  the Kirkwood function.
The first register is the ancillary (probe) qubit, the second and the third registers
are the program states. The $FT$ gate is the quantum Fourier transform, and the swap gate
acts as $S [|\Psi\rangle_1 |\Phi\rangle_2 ]= |\Phi\rangle_1|\Psi\rangle_2$.}\label{kirk}
\end{figure}

The Kirkwood function is a phase space distribution whose use is probably less common. 
It was first proposed by Kirkwood \cite{kirkwood}, and used in quantum
statistics. It displays different phase space features of a quantum state in phase space
and has the advantage of being directly linked to the matrix elements of the density 
matrix in a mixed representation. On the 
other hand the Kirkwood function  is a complex number even for hermitian operators. 
Having defined the unitary basis 
of operators $\{\frac{|q\rangle \langle p|}{\langle p|q \rangle}; q,p=0,..,N-1 \}$ in a 
Hilbert space of dim $N$, the discrete Kirkwood function of a density operator $\rho$ is defined as:
\begin{equation}
K(q,p)= \frac{1}{N} \text{Tr}\Big[\frac{|q\rangle \langle p|}{\langle p|q \rangle} \rho\Big]
\end{equation}
where $|q\rangle$ and $|p\rangle$ are position and momentum eigenstates, respectively.
The circuit that implements the measurement of the Kirkwood distribution is also based on the
scattering circuit. It can be seen in Figure \ref{kirk}, where the swap gate
acts as $S [|\Psi\rangle_1 |\Phi\rangle_2 ]= |\Phi\rangle_1 |\Psi\rangle_2$.
It is simple to show that if the program state (second and third registers) are
computational states specifying the position and momentum coordinates, this circuit allows
to measure the Kirkwood distribution at any point of the phase space.
By measuring the expectation values of $\langle \sigma_z \rangle$ and
$\langle \sigma_y \rangle$ for the ancillary qubit we obtain the Kirkwood distribution 
as
\begin{equation}
\langle\sigma_z\rangle-i\langle\sigma_y\rangle=\text{Tr}\Big[|p\rangle \langle p| \rho
|q\rangle \langle q|\Big]=K(q,p).\label{measurekirk}
\end{equation}

\section{Programmable tomography of the Husimi distribution}\label{husprog}

The Husimi function is a  well known alternative distribution in phase space, which is 
based on the use of minimum uncertainty wave packets $|\alpha\rangle$. The Husimi distribution 
is the expectation value of the density matrix in the coherent state $|\alpha\rangle$. 
This is a positive quantity that, in the continuous case, graphically displays the phase space 
contents of the state in a region of area $h$. In the discrete case the coherent 
states can also be defined \cite{Saraceno} (we provide below 
an efficient scheme for their preparation). These wave packets define an over--complete basis 
$\{|\alpha\rangle, \alpha=(q,p), q,p=0,..N-1\}$. In terms of this basis the Husimi distribution
is defined as
\begin{equation}
H(\alpha)=\frac{1}{N}\text{Tr} \Big[ |\alpha\rangle\langle \alpha| \rho \Big].
\end{equation}

\begin{figure}[ht]
\includegraphics[width=8cm]{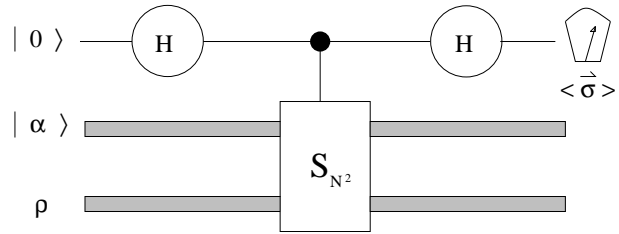}
\caption{Programmable circuit to measure the Husimi distribution function. The
first register is the ancillary (probe) qubit, the second register is the program state (a coherent
sate centered at the point of interest in phase space) and the third one is the system of
interest.}\label{hus} 
\end{figure}

The programmable circuit that implements the measurement of the Husimi distribution is also
based on the scattering circuit (in particular, it is a straightforward application of 
the ideas proposed in \cite{Ekert} and used in the experiment presented in \cite{Hendrych}). 
In Figure \ref{hus} we can see a representation of 
the algorithm, which uses an ancillary (probe) qubit, and a program register prepared in 
the state $|\alpha\rangle$ (a coherent state centered at the point where we want to evaluate 
the Husimi distribution). It is easy to show that the circuit is such that 
\begin{eqnarray}
\langle\sigma_z\rangle=\text{Tr}\Big[ |\alpha\rangle\langle \alpha| \rho\Big]=N H(\alpha).
\end{eqnarray}
The algorithm would only be useful if an efficient method can be devised to prepare the 
state $|\alpha\rangle$. We devote the next section to a complete description of this 
subroutine. 

\section{Efficient algorithm for the generation of coherent states}\label{Coh}

\subsection{Discrete coherent states}

\begin{figure}[ht]
\includegraphics[width=8cm]{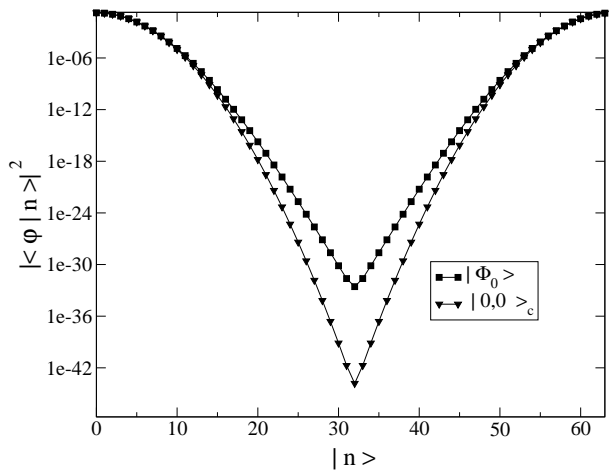}
\caption{Population (in the coordinate basis) of both the continuous (\ref{cohcont}) 
and the discrete (\ref{harper}) coherent states, centered at the origin of the phase space. 
The difference between both definitions
tends to zero in the large $N$ limit (here, $N=64$).}\label{pobla}
\end{figure}

In the continuous case, coherent states can be defined as phase space translations acting 
on  the ground state of the harmonic oscillator Hamiltonian.  In the discrete case
with periodic boundary conditions, the harmonic oscillator can be replaced by the Harper
Hamiltonian:
\begin{equation}
\Big(2-\frac{U+ U^\dag}{2}-\frac{ V+
V^\dag}{2}\Big)|\Phi_0\rangle=E_0|\Phi_0\rangle, \label{harper}
\end{equation}
which ensures the proper periodicity conditions.
Coherent states can now be defined  \cite{Saraceno} as discrete translations on $|\Phi_0\rangle$
\begin{equation}
|\alpha\rangle=T(\alpha)|\Phi_0\rangle \label{trasla},
\end{equation}
where $T(\alpha)=U^q V^p e^{\frac{i\pi}{N}pq}$ are the phase space translation operators.
An alternative definition, yielding a {\it continuous } distribution with analytic 
properties \cite{voros} is given by:
\begin{eqnarray}
|q,p\rangle_c=\sqrt[4]{\frac{2}{N}}e^{\frac{\pi}{2N}[q^2+p^2]}&\sum_{n=0}^{N-1}
\sum_{j=-\infty}^{\infty}e^{-\frac{\pi}{N}(Nj-q+n)^2}&\nonumber \\
&e^{-i\frac{2\pi}{N} p(Nj+\frac{q}{2}-n)}|n\rangle.& \label{cohcont}
\end{eqnarray}
These states are almost indistinguishable from (\ref{trasla}) as $N$ grows, and both are periodic 
wave packets occupying a minimum uncertainty area $1/N$ in phase space. Their detailed  structure,
showing the extremely small differences at small values is shown in Figure \ref{pobla}. 

\begin{figure}[ht]
\includegraphics[width=8cm]{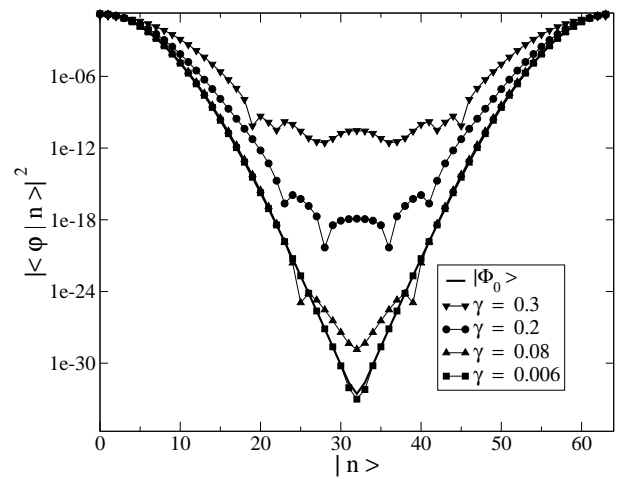}
\caption{Population (in the position basis) of the  eigenstates of the Harper Hamiltonian and 
the Harper kicked map, ${\cal U}(\gamma)$, as a function of $\gamma$. We can appreciate
that for the smallest values of $\gamma$ these states are almost identical ($N=64$).}\label{eigenevol}
\end{figure}

The classical analogue of the Harper Hamiltonian (\ref{harper}) is 
$H=\frac{1}{2}(\sin^2 \pi Q + \sin^2 \pi P)$. This gives rise to the following 
classical map equations (for a small time step $\gamma /2 \pi$) 
\begin{eqnarray}
Q'&=&Q-\gamma \sin(2\pi P) \quad\text{mod 1} \\
P'&=&P+\gamma \sin(2\pi Q') \quad\text{mod 1}.
\end{eqnarray}
For infinitesimal $\gamma$ we obtain Hamilton equations, and the conservation of energy 
leads to integrable behavior. Our strategy will be to quantize the map equation for small
$\gamma$ as a way to obtain a unitary operator with an eigenstate very close to $|\Phi_0\rangle$. 
The map  belongs to the well known family of the kicked maps and the unitary operator corresponding
to its quantization is obtained as a product of two operators representing a potential and a kinetic 
kick.  These two operators are respectively diagonal in position and momentum basis. They can 
be efficiently implemented by means of a quantum network consisting of two controlled phases
interposed by the Fourier transform:
\begin{equation}
{\cal U} (\gamma)=M [FT]^{\dagger} K [FT],
\label{evolmapa}
\end{equation}
where $FT$ is the $N$-dimensional Fourier transform and the operators
$M$ y $K$ represent the potential and kinetic kicks respectively:
\begin{eqnarray}
M|q\rangle&=&e^{-i\gamma N \cos(\frac{2\pi}{N}{q})}|q\rangle\\
K|p\rangle&=&e^{-i\gamma N \cos(\frac{2\pi}{N}{p})}|p\rangle.
\end{eqnarray}
The crucial feature of this unitary operator is that as $\gamma\ll 1$ its eigenstates
become those of  Harper Hamiltonian. Hence, ${\cal U}(\gamma)$ has a coherent
state as one of its eigenstates for small values of $\gamma$. In Figure \ref{eigenevol} we
show how the eigenstate of the map (\ref{evolmapa}) converges to the ground state $|\Phi_0\rangle$ 
as $\gamma \to 0$. 

From the above discussion is clear that what we need is a method to efficiently prepare 
an eigenstate of the unitary operator (\ref{evolmapa}). For this, we will use the 
well known phase estimation algorithm  \cite{Nielsen,Cleve,Lloyd} to filter
an initial state which is  approximately localized near the phase space origin.

\subsection{Algorithm for phase estimation}

\begin{figure}[ht]
\includegraphics[width=8cm]{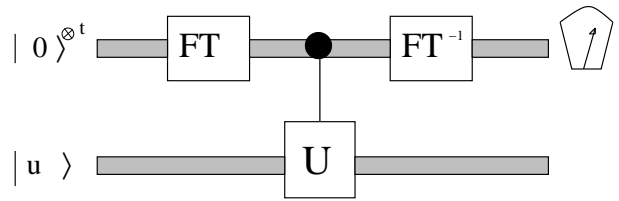}
\caption{The circuit for the phase estimation algorithm. When the state of the lower register is an
eigenstate of $U$, the measurement of the upper register reveals an $t$-bit approximation
to the corresponding eigenphase.}\label{phase}
\end{figure}

As the phase estimation algorithm \cite{Nielsen,Cleve,Lloyd}, is an essential part
of our construction we briefly review its operation. The circuit is reproduced in Figure \ref{phase}.
Its operation in the phase estimation mode requires that an eigenstate of ${\cal U}(\gamma)$ be 
supplied to the lower register. Then a measurement performed in the upper registers yields a 
rational approximation to the eigenphase, which improves as the size of the upper register increases.
Here we are more interested in the use of the circuit as a filter in which case a state
approximating an eigenstate is fed to the lower register.  If this state is expanded as 
$|\Psi\rangle = \sum_\alpha c_\alpha |u_\alpha\rangle$ the measurement in the upper register yields
a distribution of phases with probabilities proportional to $|c_\alpha|^2$.
Furthermore, if the number of qubits in the first register is such that
the value of the phase can be exactly determined, then the final state
of the second register is the eigenstate corresponding to that phase. Thus, it is clear that
if the initial state $|\Psi\rangle$ is near $|\Phi_0\rangle$, the application of this 
circuit would provide us with the desired state with high probability (see below).   

However, the typical situation is when the number of qubits in the first register only 
allows us to make an approximation to the real value of the phase. Thus, if the system 
has $n$ qubits and the number of qubits in the first register is
\begin{equation}
t=n+\log_2\big[2+\frac{1}{2\epsilon}\big], \label{time}
\end{equation}
the phase estimation algorithm gives an approximation to the phase corresponding to the 
eigenstate $|u\rangle$, with probability bounded by 
\begin{equation}
p_u=|c_u|^2(1-\epsilon). \label{prob}
\end{equation}
This clearly differs from the ideal case, since now we do not obtain the exact value of
the phase. Therefore the above probability corresponds to the measurement of an integer $k$
 such that $\frac{k}{T}$ is the best $t$-bit estimate to $\varphi$ (with an error bounded by $|2^{t-n}-1|$).  
After the measurement, the state of the second register will be a linear combination of all the eigenstates
of $U$ that is close to the corresponding eigenstate (see below).

\subsection{Algorithm for the generation of coherent states}

As mentioned above, the algorithm for the preparation of coherent states consists of 
the application of the phase estimation algorithm to filter an initial state (which should
be itself a well localized state near the origin of phase space). We will now 
describe in detail the two necessary ingredients for the efficient
implementation of the algorithm: i) the preparation of the initial state  
and, ii) the efficient implementation of the ctrl-${\cal U}^j(\gamma)$ gates, required for 
the phase estimation to be applicable.

\subsubsection{Initial state preparation}

\begin{figure}[ht]
\includegraphics[angle=-90,width=6cm]{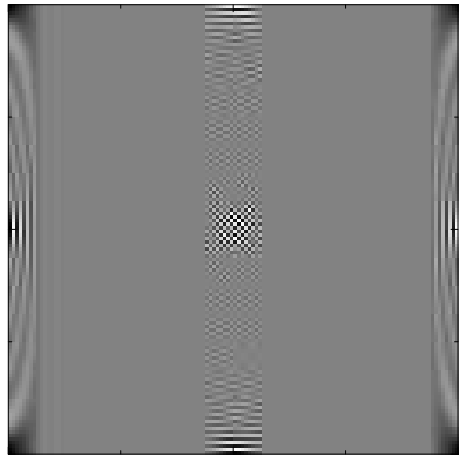}
\caption{The Wigner function of a square state. The state is located in position 
and partially localized in momentum (N=64). Horizontal (vertical) axis corresponds
to position (momentum) basis. The color convention is such that positive (negative) values 
of the Wigner function correspond to black (white) regions. }\label{square} 
\end{figure}

This is indeed the simplest part of the algorithm. In fact, an easily preparable 
candidate for the initial state is what we could denote as a ``square state'', 
defined as an equally weighted superposition of the first $\sqrt{N}$ states 
of the computational basis:
\begin{equation}
|\Psi_0\rangle=\frac{1}{w^{1/2}}\sum_{q=0}^{w-1}
|\frac{w}{2} -q\rangle \quad \text{mod $N$},
\end{equation}
where $N$ is the Hilbert space dimension and $w=[\sqrt{N}]$.  This state is strictly localized  
in position in a region of width  $w$ around the origin but, because of diffraction, 
is only partially localized in momentum. This can be seen by analyzing its 
Wigner representation shown in Figure \ref{square}.
The important features of this state are that it has strong overlap with a coherent state localized
at the origin of phase space (which is our target state). In the limit of large $N$, the overlap
tends to a value of $0.94$. Also, it is simple to show that the square state 
can be efficiently prepared: Starting from $|0\rangle ^{\otimes n}$ we simply need to apply
Hadamard gates to the $n/2$ least significant qubits we obtain a state centered at the phase
space point $(w/2,0)$. Centering the state at the origin requires a shift, which can be implemented 
efficiently. 

\subsubsection{Efficient implementation of the ctrl-${\cal U}^j(\gamma)$ gates}

The phase estimation algorithm requires that the powers of the operator ${\cal U}(\gamma)$  
be implemented efficiently. This is in general not the case. However, we can get around this 
problem by using the fact that the Harper map has a well defined semi-classical limit. This 
allows us to perform its iteration by using a reliable semi-classical approximation. 
Thus, for  $\gamma\ll 1$ (and for $N$ large) the powers needed can be approximated as follows: 
\begin{equation}
{\cal U}^{2^t}(\gamma)\simeq  {\cal U}(2^t\gamma). \label{semiclassical}
\end{equation}
This approximation also relies on the assumption that the initial state is localized in a region
of the phase space where the map is regular (this is satisfied by the square state).
If this approximation is valid, then the powers of the unitary operator can be 
implemented efficiently using the same quantum networks required to implement the 
operator itself (to implement a power of $U$ we simply use a different parameter $\gamma$). 
This approach has a clear limitation: Each power increases the value of $\gamma$ and 
for large values of $\gamma$ the map ceases to be integrable. In such case, its  
spectral properties become very different from those of the Harper Hamiltonian. 
Our goal then is to propagate the map for a time long enough to resolve the 
ground state from its neighbors without violating the approximation (\ref{semiclassical}).
These two issues (evolving accurately and resolving the spectrum) should be studied
jointly. But to make our presentation clear we can first analyze them separately.

\begin{figure}
\includegraphics[width=8cm]{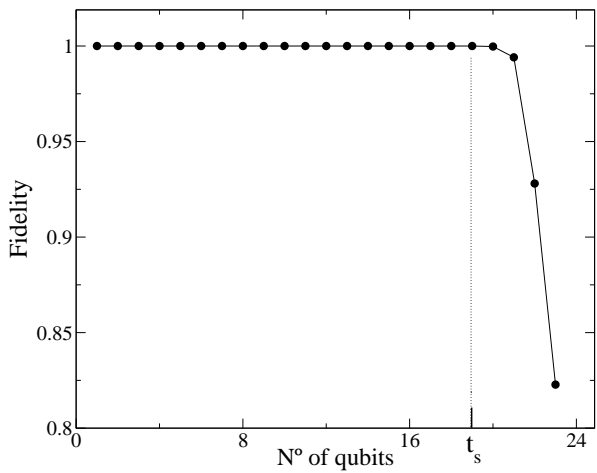}
\caption{The fidelity (overlap between the state evolved with the exact and approximated
unitary operators) as a function of the number of qubits in the first register. $2^{t_s}$ 
defines the maximum time for which (\ref{semiclassical}) is valid.} \label{fidelity}
\end{figure}

To  examine the accuracy of the approximate evolution we can study  
the fidelity, defined as the absolute value of the overlap between the square state propagated with the exact 
and approximate evolution (i.e. $F=\sqrt{|\langle\psi_{exact}|\psi_{approx}\rangle|}$). In Figure \ref{fidelity}
we plot this as a function of the number of qubits in the first register.
We find that the fidelity remains close to unity
up to a sharply defined time $T_s$  $(T_s=2^{t_s})$, after which it drops abruptly. 
This time, $T_s$, defines the allowed number of iterations of the unitary compatible
with a given accuracy. Therefore, it sets a bound to the maximum number of qubits
we can include in the first register of the phase estimation algorithm for the 
semi-classical approximation to remain valid.

\begin{figure}[ht]
\includegraphics[width=8cm]{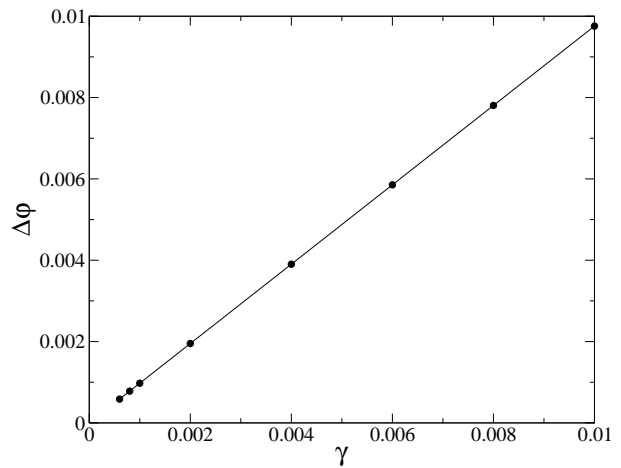}
\caption{Dependence of the phase difference $\Delta \varphi$ with $\gamma$. A linear
dependence is found for values of the Hilbert space dimension  
$N$ such that $N\gamma<0.6$.} \label{phi_gamma}
\end{figure}

To analyze the resolution required to resolve the spectrum of the unitary operator we
should consider the minimum difference between neighboring eigenphases of ${\cal U}(\gamma)$ 
(denoted as $\Delta \varphi$). This quantity determines the 
minimum number of qubits in the first register, needed to resolve the
ground state. Thus, for this purpose, we would need a number of qubits $n$ which should
be at least equal to 
\begin{equation}
n=\log_2(\frac{1}{\Delta\varphi}). \label{cota}
\end{equation}
An important question is how does this number scales with the dimension of the 
Hilbert space of the system. This can be determined by analyzing the dependence of 
$\Delta \varphi$ with $\gamma$ and $N$. This is done 
in Figure \ref{phi_gamma} where we show that the phase difference has a linear 
dependence with $\gamma$, at least for values of $N$ and $\gamma$ such 
that $N\gamma<0.6$. Therefore, the number of qubits required to resolve the spectrum
will scale logarithmically with the dimensionality of the system's Hilbert space. 

\begin{figure}[ht]
\includegraphics[width=8cm]{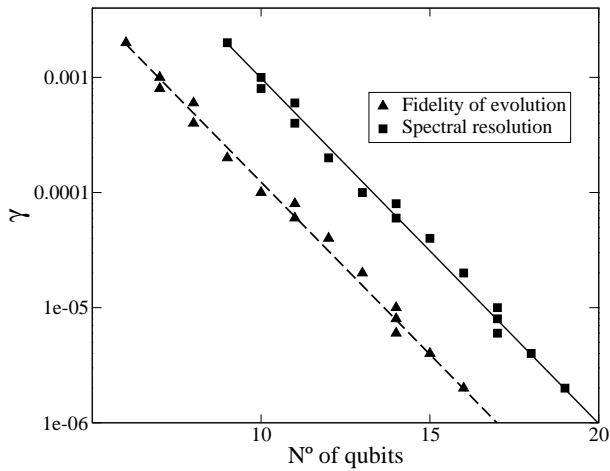}
\caption{Dependence of the parameter $\gamma$ as a function of the number of qubits 
in the first register. The dashed line represents the condition of high 
fidelity in the evolution (the allowed parameter region lies below this curve). The 
solid line corresponds to the condition of good spectral resolution (the allowed
parameter region lies above that curve). The two lines are shifted from 
each other by about three qubits.}\label{evol_pres}
\end{figure}

As mentioned above, the behavior of the fidelity as a function of the parameter 
$\gamma$ should be analyzed jointly with the minimum number of qubits needed in 
the first register, to achieve the required spectral resolution 
according to eq. (\ref{cota}). In Figure \ref{evol_pres} we display the curves 
coming from each of these requirements. To achieve high enough
fidelity the value of $\gamma$ and of the number of qubits in the first register
must be below the lower curve. On the one hand to achieve the required spectral resolution one needs 
the value of $\gamma$ and the number of qubits to
lie above the second curve. This seem to be a problem, but can be easily solved
taking into account the following observations. Thus, we notice that: i) the two lines 
are approximately parallel (for all values of $N$, as long as $N\gamma<0.6$) and 
ii) the two parallel lines are simply shifted away from each other by about 
three qubits. Hence, the semi-classical approximation can be used up to a power
of ${\cal U}(\gamma)$ given by ${\cal U}(\gamma)^{2^{n-4}}$
with $n=\log_2(\frac{1}{\Delta \varphi})$. To iterate the map further, as required
to achieve enough spectral resolution we can apply the remaining powers of $U(\gamma)$
as products of its precedents (which were implemented using the semi-classical approximation).
As the distance between curves is fixed, we would only 
need to use this trick a fixed ($N$--independent) number of times. Doing this, it
is possible to achieve the required accuracy and spectral resolution simultaneously. 

It is worth mentioning that the precision needed to resolve the ground state is not 
the only condition that imposes a lower bound on the number of qubits of the first register. 
Thus, in principle if we want to get the desired coherent state with a reasonable probability
we need the number of qubits to obey the relation fixed by equation (\ref{prob}). 
Suppose that we impose a value of $\epsilon=1/4$, which corresponds to 
a probability for preparing the right coherent state of about $p_0=0.70$ (this is 
computed taking into account that the initial square state is such that $|c_0|^2=$0.94).
For this value of $\epsilon$, equation (\ref{prob}) implies that the first register
should have at least two more qubits than the system's register. 
This is a lower bound for the dimension of the first register.

\subsubsection{Final remarks on the preparation of coherent states}

In summary, the algorithm to prepare a coherent state consists of
the following steps: i) preparation of a ``square'' state, ii) selection of the 
parameter $\gamma$ for the evolution operator and the corresponding determination of 
the number of qubits to be used in the first register of the phase estimation 
algorithm, iii) run the phase estimation algorithm efficiently implementing the 
powers of the operator ${\cal U}(\gamma)$ in an approximate way,
iv) from the peaked distribution of results for the phase, we discover the one 
associated with the coherent state,  when this phase is detected the desired 
coherent state has been prepared in the system's register, 
v) after obtaining a coherent state centered at the origin, one can 
translate it to any point of the phase space using phase space displacement operator, 
which can be efficiently implemented as in the circuit of 
Figure \ref{wigner}.

\begin{figure}[ht]
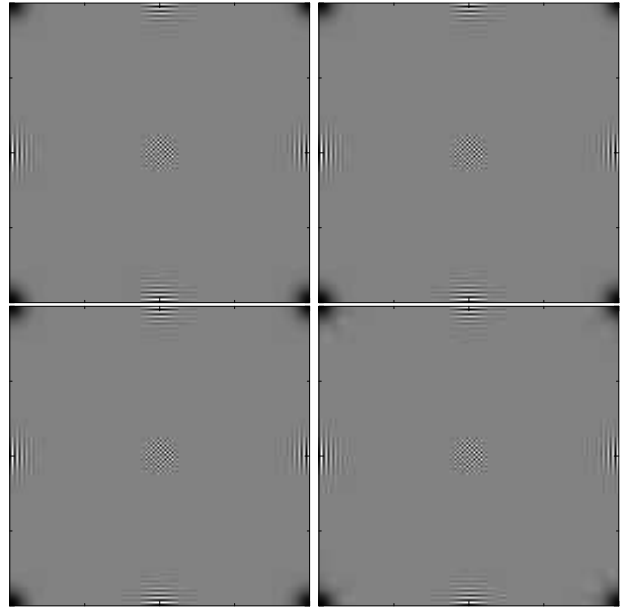

\includegraphics[angle=-90,width=4cm]{fig12a.eps}
\includegraphics[angle=-90,width=4cm]{fig12b.eps}
\includegraphics[angle=-90,width=4cm]{fig12c.eps}
\includegraphics[angle=-90,width=4cm]{fig12d.eps}
\caption{Wigner function for four of the quantum states of the system's register that are 
generated by the phase estimation algorithm when the detected value of the first register 
falls in the peak of the probability distribution. Horizontal (vertical) axis corresponds
to position (momentum) basis. Labeling these states from left to right and from top to bottom as a), b), c) and
d), we can see in Figure \ref{filter2} their representation in the computational basis.}\label{filter1} 
\end{figure}

\begin{figure}[ht]
\includegraphics[width=8cm]{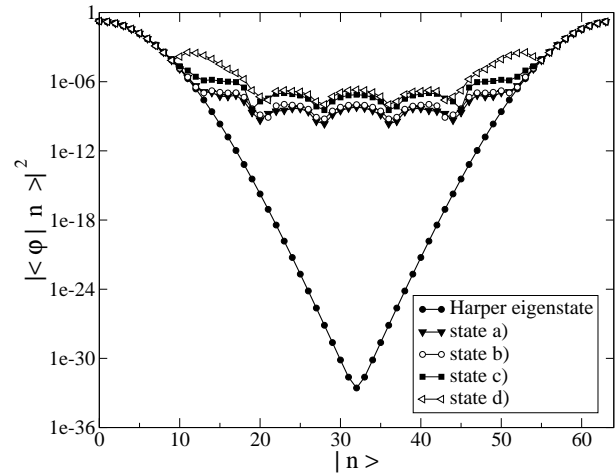}
\caption{Probability distribution in the computational basis for four of the 
quantum states of the system's register that are prepared when the 
detected value of the first register falls in the peak of the distribution.
The Wigner function of such states is shown in Figure \ref{filter1}.}\label{filter2}
\end{figure}

\begin{figure}[ht]
\includegraphics[width=8cm]{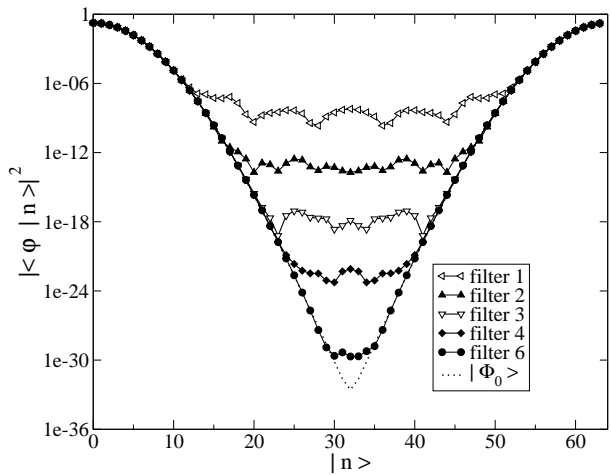}
\caption{Probability distribution in the computational basis for the states
produced after different number of iterations of the filtering algorithm. Each
iteration improves the quality of the state, which becomes closer to a 
true coherent state.} \label{filter3}
\end{figure}

To see the algorithm in action we performed a few numerical simulations.  In Figures 
\ref{filter1} we show the Wigner function of four of the quantum states of the system's 
register that fall under the peak of the probability distribution for the first 
register (we used $N=64$ and showed in Figure \ref{filter2} the probability distribution 
for the same states in the computational basis). It is clear that any of such states
is a good approximation to a coherent state (the fact that this stage of the algorithm
works as a filter can be appreciated by comparing the initial square state shown in 
Figure \ref{square} and those shown in Figure \ref{filter1}).  
In Figure \ref{filter3} we show how the approximation can be improved even further by
various means. In fact, we could include several stages of filtering each one
of which would considerably improve the quality of the final state. 

\section{Conclusions}\label{Concl}
In this paper we presented various algorithms to evaluate several phase space 
distributions of arbitrary states, These methods allow, in principle, to perform
phase space tomography in an efficient manner. 
The efficiency of the circuits is based on the fact that operations such as phase 
space translations, reflections and the Fourier transform are efficiently implementable. 
For the case of Wigner and Kirkwood distributions, the efficiency is solely based
on this fact (however, it is worth pointing out that, contrary to what happens with the 
Wigner and Husimi distributions, the evaluation of a typical value
of the Kirkwood distribution of a pure state would require exponential precision due to 
the factor of $N$ absent in (\ref{measurekirk}) as compared with (\ref{measurewig})).  
The evaluation of the Husimi distribution requires the use of a subroutine preparing 
coherent states. We presented a method achieving this goal, which consists of a variation
of the phase estimation algorithm with an appropriately chosen initial state. 
In this case the efficiency requires not only a good guess for the initial state (which is 
indeed easily done) but also the possibility of efficiently implementing powers of 
the unitary map whose eigenstate is close to a coherent state. In our case, this can be done
by using a semi-classical approximation for this operator.

\begin{acknowledgments}
This work was partially supported with grants from Ubacyt, Anpcyt  
03-9000, Conicet and Fundaci\'on Antorchas. JPP and AJR were also partially supported by
a grant from NSA. 

\end{acknowledgments}

\end{document}